# An Extremely Flexible, Energy, and Spectral Effective Green PHY-MAC for Profitable Ubiquitous Rural and Remote 5G/B5G IoT/M2M Communications

Alexander Markhasin(✉)

Siberian State University of Telecommunications, Novosibirsk, Russia
almar@risp.ru

**Abstract.** In this paper, the fundamental PHY-MAC throughput limits and extremum of the energy, power, spectral efficiency invariant criteria are proved. The invariant criteria are constructed relying on Shannon's *m*-ary digital channel capacity which a rich palette of the technically interpreted PHY-MACs parameters consider. Therefore, the invariant criteria as very suitable for research and design of an 5G extremely performance problems are found. The PHY-MACs smart distributed control techniques which able implements "on-the-fly" the limits close and invariant criterion optimization or trade-off is proposed. Such PHY-MAC's smart control techniques represent a key disruptive technologies meet the 5G/B5G network challenges.

## 1 Introduction

Unacceptably high investments are required into deployment of the optic core infrastructure for ubiquitous wide covering of sparsely populated rural, remote, and difficult for access (RRD) areas using the recent (4G) and also forthcoming (5G) broadband radio access (RAN) centralized techniques, characterized by short cells ranges, because their profitability boundary exceeds a several hundred residents per square kilometer. Furthermore, the unprecedented requirements and new features of the forthcoming Internet of Things (IoT), machine-to-machine (M2M), smart city, and also many other machine type IT-systems lead to a breakthrough in designing extremely intensive technologies for future 5G/B5G wireless systems which will be able to reach in real time the performance extremums, trade-off optimums and fundamental limits [1–3]. Recently, a number of 5G extremely intensive solutions were proposed which are suitable mainly for well-urbanized areas: ultra-dense networks [4], massive MIMO [5] and M2M for smart city [6], disruptive 5G PHY technology [2].

For weakly urbanized areas, we offer extremely effective green [7] techniques as an approach for ubiquitous profitable covering by 5G/B5G IoT/M2M/H2H multifunctional communications of the RRD territories [8, 9]. Practically, the necessary and sufficient conditions for RRD areas profitability border overcoming envisages three extremal RRDs networking performances': (i) the hyper long range hypercells' radically distributed cost-effective Multifunctional Hyperbus Architecture (MFHBA) [8, 9] and MFHBA mission-critical convergent techniques – (ii) the extremely energy-effective green PHY [10], and (iii) the supreme throughput capacity multifunctional MAC [11]. Convergent PHY-MACs shall close the Shannon's fundamental limits or extremums using the multifunctional optimal "on-the-fly" control techniques [8, 12].

Recently [3, 7], the spectral (SE) and energy (EE) efficiency criteria are expressed usually through the Shannon's capacity of the continuous channels with additive white Gaussian noise (AWGN) [13] which, in principle, allow to study only the PHY potential efficiency values depending directly on three spectral-power basic parameters, i.e., bandwidth $\Delta F_s$, signal power $P_s$ and AWGN noise power $P_n$. In [10, 14] so called invariant criteria of spectral (ICSE), power (ICPE) and energy ICEE) efficiency were first introduced for orthogonal spread spectrum m-ary signals. The invariant criteria are constructed relying on Shannon's m-ary digital channel capacity which considers a rich palette of the technically interpreted PHY-MAC parameters. Therefore, the invariant criteria were found very suitable for research and design of an 5G extremely performance problems.

In this paper, we generalize and develop the results of our above cited researches of the RRD radically distributed multifunctional device-centric MFHBA architecture, fundamental RRD PHY and information-theoretic RRD MAC limits and extremums focused on the 5G/B5G extremely performance issues and also PHY-MAC multifunctional optimal control techniques which meet green profitable ubiquitous rural and remote 5G/B5G IoT/M2M/H2H communications. The conceptual vision of the green profitable ubiquitous RRD 5G/B5G IoT/M2M/H2H architecture is also presented.

## 2 Extremely Green and Cost-Effective Ubiquitous RRD 5G PHY

### 2.1 Vision of Extremely Green and Effective 5G PHY for RRD Areas

As stated above, the widespread cell range of the recent (4G) and forthcoming (5G) generations of radio access technologies (RAN) does not exceed few kilometers. The sparsely populated RRD areas differ by low density up to a few tens residents. Hence, the really indispensable approach for overcoming the 5G RRDs economical barrier lead to extremely increasing of the number of the effectual subscribers, i.e., to increasing of the air interface range of broadband hypercells by several ten times through approaching the fundamental Shannon limits of spectral (SE) and power (PE) efficiency. Usually [3, 7], the SE and PE efficiency criteria are expressed through the Shannon's capacity of the continuous channels with additive AWGN noise [13]

$$C = \Delta F_s \log_2(1 + P_s/P_n), \quad (1)$$

where $\Delta F_s$ is bandwidth, $P_s$ – signal power, $P_n$ – noise power, $P_n = \Delta F_s N_0$, $N_0$ – signal-sided spectral power noise density, in Watt-per-Hertz. In the channel output, or receiver input, power characteristic $P_s/P_n$ is called the signal-to-noise-ratio (SNR).

However, the continuous channel throughput capacity (1) allow to study only the potential efficiency PHY values depending directly on three spectral-energy basic parameters. So called invariant criteria of spectral, power and energy efficiency [10] allow to solve an optimization or trade-off problem depending on the set of real conditions and parameters of the radio channel, methods of signal coding, formation, modulation, transmitting, receiving, processing, decoding, etc. Two invariant efficiency criteria were first introduced for the wireless physical layer with orthogonal spread spectrum $m$-ary signals in [14] basing on Shannon's $m$-ary digital channel capacity. As in [10], let us introduce an invariant efficiency criterion for modern 5G PHY relying on SINR [15] approaches. The invariant criterion for spectral efficiency (ICSE) was introduced as the digital channel Shannon capacity per Hertz ((bit/sec)/Hz):

$$c_F(m,g,B_s) = C_m(g,B_s)/(B_s/2) \tag{2}$$

where $g$ is channel-side, or receiver input, mean square signal power invariant variable expressed via signal-to-interference plus noise ratio (SINR) [15],

$$g^2 = P_s/(P_i + P_n) \tag{3}$$

$P_s$, $P_i$, $P_n$ are, respectively, signal, interference, and noise powers, $B_s$ is frequency-time invariant variable named as signal's base, $B_s = 2\Delta F_s T_s$, $T_s$ is $m$-ary signal duration. Further, $C_m(g,B_s)$ is $m$-ary digital channel Shannon capacity in bit-per-symbol [10],

$$\begin{aligned} C_m(g,B_s) = \log_2 m + [1 - p_m(g,B_s)] \log_2[1 - p_m(g,B_s)] \\ + p_m(g,B_s) \log_2[p_m(g,B_s)/(m-1)], \end{aligned} \tag{4}$$

where $p_m(g,B_s)$ is $m$-ary symbol's error probability (SER) [15] defined through invariant variable $h(g,B_s) = g\sqrt{B_s/2}$ expressed, in turn, through receiver's output ratio signal energy per symbol to signal-sided spectral power additional Gaussian interference plus noise density $N_{0_{in}} = N_{0_i} + N_{0_n}$, i.e., energies SINR, or ESINR [10]:

$$h^2 = E_s/N_{0_{in}} = P_s B_s/[2(P_i + P_n)] = g^2 B_s/2\,. \tag{5}$$

An invariant criterion for power efficiency (ICPE) was introduced as the signal-to-interference plus noise ratio (SINR) per $m$-ary digital channel Shannon capacity per Hertz: SINR/[(bit/sec)/Hz] [10]:

$$w(m,g,B_s) = g^2/c_F(m,g,B_s). \tag{6}$$

One can express a power efficiency criterion (6) through various measure units: dBm per (bit/sec)/Hz, Watt per (bit/sec)/Hz, and also convert it to energy efficiency invariant criterion (ICEE) in Joule per (bit/sec)/Hz. Moreover, through invariant criterion for power efficiency (6) one can express the invariant criteria for cover efficiency (ICCE) in Watt/(bit/sec)/Hz/square km, i.e., ICPE per area covering by cell radius $R_c(m,g,B_s)$ and also invariant criterion for investment (cost) efficiency (ICIE) through CAPEX calculated as some invariant function $F_I[w(m,g,B_s)]$ divided into area covering $\pi R_c^2(m,g,B_s)$.

Based on the introduced invariant criteria, we can formulate the following RRD-aimed breakthrough qualities and techniques capable to implement the perfect green 5G PHY for hyperrange space/wireless mediums corresponding to rural ubiquitous IoT/M2M/H2H 5G communications:

- design an advanced set of the orthogonal broadband *m*-ary OFDM-CDMA like waveforms corresponding to a perfect green 5G PHY for hyperrange space/wireless mediums which are well adapted to cognitive interference-robust "on-the-fly" control and approaching the trade-off extremums or fundamental limits of the spectral/power/energy/economics efficiency criteria [10];
- refine the green 5G PHY disruptive approaches for the potentially reachable energy-saving techniques of hyperrange rural area cost-effective covering;
- increase the channel-side ratio SINR (3) in pure ecological way of improvement both the denominator (reduce an interference [16]), and the numerator (smarter increase a beamforming and antenna gain, as Friis models), close to the fundamental limits without the rise of transmitter power;
- reaching continuously the fundamental minimum [10] power consumption criterion ICPE representing an imperative law for smart green PHY optimization and trade-off problems;
- as in [3], developing the profitability-power-efficiency-aimed fundamental trade-offs for rural green 5G networks in practical invariant variables notions.

## 2.2 Fundamental Limits and Extremums of 5G PHY

The fact that the value of invariant function $F(x_1, x_2\ldots)$ does not change by substitution instead of every $x_i$ argument's his $x_i^*(x_1, x_2, \ldots)$ invariant maps may be suitable for universal appropriateness research all measures: the information, the power, the covering, and the investment (i.e., profitability) measures. Let us denote by $U$ the set of possible values of the invariant parameters $(m, g, B_s)$. In a specific optimization problem some invariant variables are free and other parameters are fixed. We denote the set of possible values of the free variables by $V$, $V \in U$. Next we can formulate a set of general optimization problems [10].

**Power Efficiency Optimization Problem.** For ICPE (6), we can formulate the general optimization problem

$$w(m, g, B_s) \to \min, \tag{7}$$

where a free variable belongs to $V$. It is necessary to bind the problem (7) with the constraint on the least permissible value $[c_F]_{\min}$ of ICSE

$$c_F(m, g, B_s) \geq [c_F]_{\min} \tag{8}$$

and, possibly, the constraints on the permissible values of cover efficiency ICCE and investment efficiency ICIE, i.e. profitability. The example of the numerical analysis [10] of ICPE optimization problem is shown in Fig. 1.

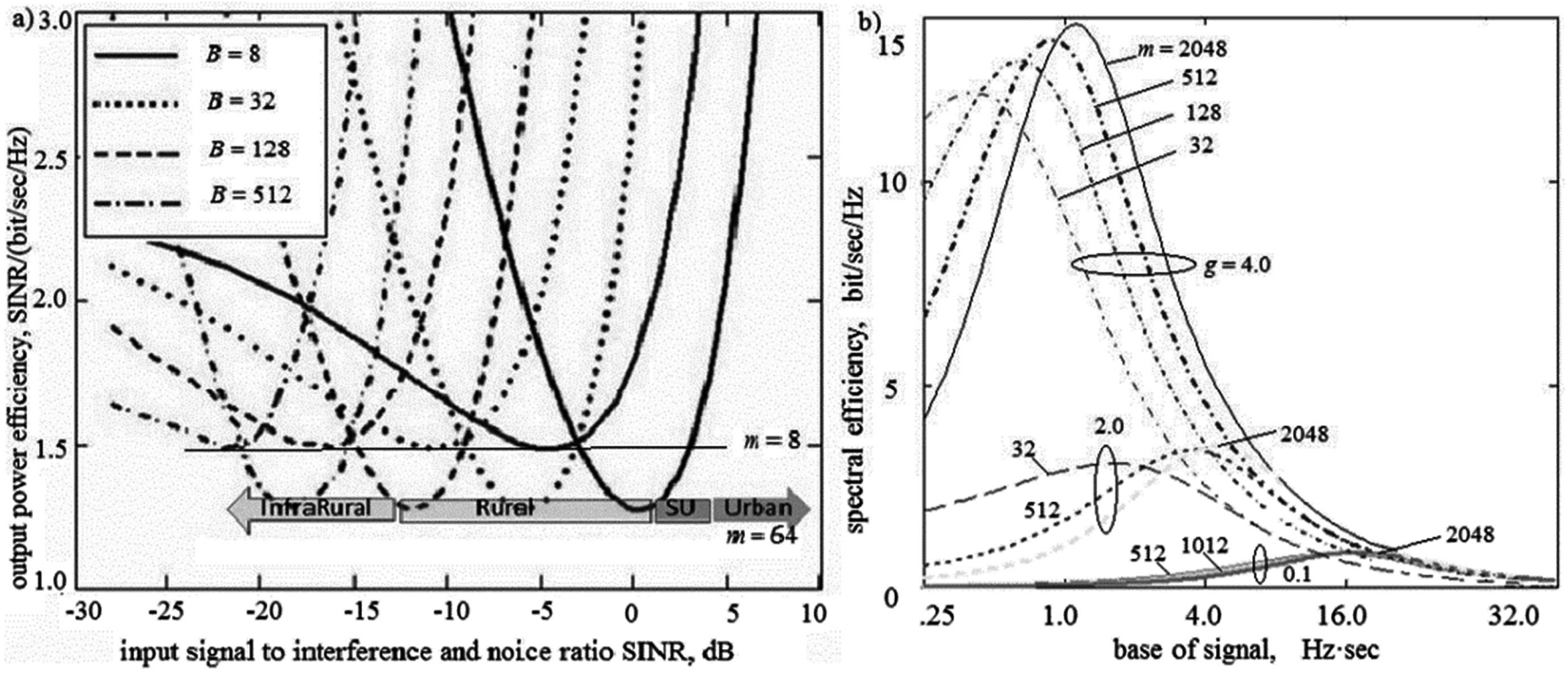

**Fig. 1.** The graphs of general optimization problems of green invariant efficiency criterion: (a) close fundamental minimum limit (infimum) of power efficiency ICPE [10]; (b) close upper limits of spectral efficiency ICSE (calculation: T. Pereverzina, D. Shatsky).

Studying the Fig. 1a and [14], we can formulate a fundamental power-consumption

**Statement 1:** The minimal specific power consumption (6) $w_{\min}(m, g, B_s)$ per (bit/sec)/Hz for fixed alphabet size $m$, and free $g$ and $B_s$ for both Gaussian noise and interference is a universal power constant which depends neither on the signal base $B_s$ nor on SINR (3).

Let $w^*_{\min}(m, g^*, B^*_s)$ be some minimum point on graph of Fig. 1a which was expressed in SINR-per-(bit/Hz)/sec. We can express this minimum value in Joule-per-(bit/Hz)/sec using the invariant relationship $w^*_{Jc}(m, g^*, B^*_s) = w^*_{\min}(m, g^*, B^*_s) \times N^*_{0_{in}} B^*_s/2$, where $N_{0_{in}}$ is the value realized in minimum point of both Gaussian interference and signal-sided noise spectral power density as in (5), $N_{0_{in}} = N_{0_i} + N_{0_n}$, in Watt-per-Hertz. Moreover, we can express this minimum value in Joule-per-bit $w^*_{Jb}(m, g^*, B^*_s) = w^*_{Jc}(m, g^*, B^*_s) \times B^*_s/2$.

**Spectral Efficiency Optimization Problem.** For ICSE (2), we can formulate the general optimization problem [10]:

$$c_F(m, g, B_s) \to \max \tag{9}$$

with respect to free variables belonging to the subset $V$, $V \in U$. The problem (9) expediently be bound with constraints on the infimum value of ICPE criterion

$$w(m, g, B_s) = \text{const}(m) \equiv w_{\inf}(m, g^*, B^*_s) + o(w), \tag{10}$$

where *o(w)* is Landau Small Symbol. The Eqs. (9) and (10) determine the fundamental extremum of the invariant power efficiency ICPE (6) as in

**Statement 2:** The fundamental local maximum, or conditional supremum, of the invariant spectral efficiency (2) under the condition of minimal power consumption, or conditional infimum (10), equals the solution of the problem (9).

Figure 1b shows three subsets of ICSE spectral efficiency optimization graphs accordingly to three fixed values of SINR invariant variable (g = 0,5/1,0/2,0) and different signal alphabet sizes m with dependency from signal base $B_s$ variations. In fact, the given series of SINR express the changes of channel quality from very poor up to average. Observing the numerical optimization graphs according to the given SINR series and correlating the signal complexity with signal base $B_s$ values presented on Fig. 1b graphs and [14], we can state the following fundamental ICSE.

**Statement 3:** Optimal signals according to the spectral efficiency criterion (2) should be more complicated as the quality of SINR of the channel is worse, and, on the contrary, these signals should be easier when the quality of the SINR is better.

The above formulated statements lead to extremely green strategies of minimal power consumption and energy saving for both the ultra-dense urban and also the ultra-covering or extremely cost-effective rural optimization problems.

### 2.3 Fundamental Limits of *m*-ary Orthogonal Signal Interference

As shown in [16], the errors of inaccurate fulfillment of conditions of mutual signals orthogonality inevitably generate the intra-cell and inter-cell interference that determines the available value of the SINR ratio. The SINR value, in turn, limits the capacity of cellular cell. In [16], an advanced calculation method of the CDMA network capacity is offered, which allows to consider the dependences "SINR versus not strict orthogonality errors" directly through the orthogonal signals autocorrelation and mutual correlation functions. It is shown, that it is possible to raise many times the SINR or network capacity by reduction of the signal orthogonality errors. The statements concerning fundamental limits for interference power are proved [16]:

**Statement 4:** If the errors E caused of the not-strictly orthogonality of the intra-cell *m*-ary orthogonal signals ensembles can be reduced as wished, then the power P(E) of intracell interference can be asymptotically decreased up to as much as small values:

$$\lim_{\mathrm{E}\to 0} P_{\mathrm{intra-cell}}(\mathrm{E}) = \lim_{\mathrm{E}\to 0} \sum_{j=0, j\neq i}^{n-1} \frac{1}{T} \sqrt{\mathrm{M}[K_{ji}^2(t, \mathrm{E}_j)]} = 0, \tag{11}$$

where $K_{ji}^2(t, E_j)$ is the intra-cell mutual correlation function.

Figure 2 explains the impact of the reduction of the signal orthogonality errors on the raise many times of the SINR or the network capacity.

**Statement 5:** If the errors E caused by the not-strict orthogonality of the inter-cell *m*-ary orthogonal signals ensembles can be reduced as wished, then the power P(E) of inter-cell interference can be asymptotically decreased to small values of an order of Landau Big Symbol $0(\mathrm{M}[a], 1/\sqrt{n})$:

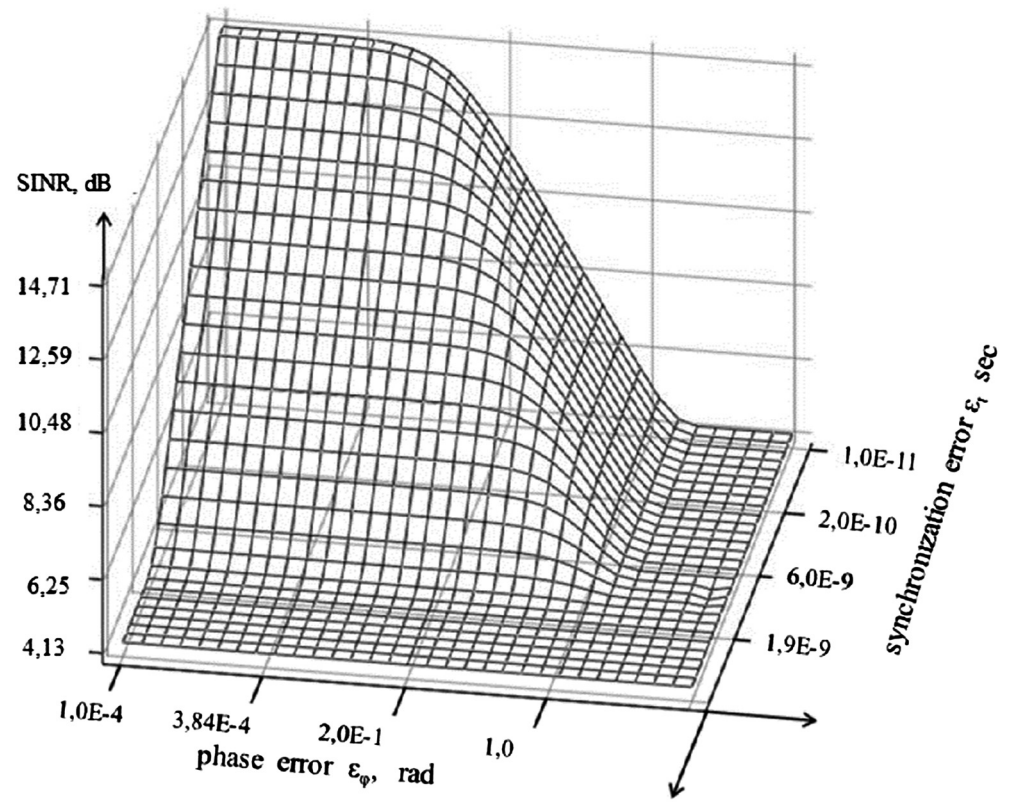

**Fig. 2.** 3D graphs SINR versus standard deviations of the synchronization $\varepsilon_t$ and phase error $\varepsilon_\phi$ by thermal noise −113.101 dB [16].

$$\lim_{\mathrm{E}\to 0} P_{\text{inter-cell}} = \lim_{\mathrm{E}\to 0} \sum\nolimits_{J\in G\backslash I} \sum_{J\in G\backslash I} \sum\nolimits_{j_J}^{n-1} M[a_{j_J}]/T\times \sqrt{\mathrm{M}[K^2_{j_J i}(t,\mathrm{E}_{j_J})|_{j_J\in J, i\in I}]} = 0(\mathrm{M}[a], 1\sqrt{n}), \tag{12}$$

where $K^2_{j_J i}(t, E_{j_J})$ is the inter-cell mutual correlation function, $\mathrm{M}[a]$ is the weighted average of the space path loss indexes $a_{j_J}$, $n$ – the degree of the generating M-sequence polynomial.

# 3 Extremely Flexible and QoS-Guaranteed Distributed Multifunctional RRD 5G MAC

## 3.1 Vision of the Distributed Multifunctional Perfect 5G MAC

Assume that at some time $t$ some quantity $N_t$ of machine type network's $i^{th}$ devices'/users' which defined by $k^{th}$ data service classes, $G_{ikt}$ input traffics intensities, $S_{ikt}$ output traffic intensities, total traffic $G_t = \sum_{i,k} G_{ikt} \leq C_{MAC}$, where $C_{MAC} = \max\limits_{\{G_t, t\}} \sum_{i,k} S_{ikt}(G_t)$ is MAC useful throughput, is active. Let we denote further by $X_{ikt}$ the really values of service parameters by $[X_{ikt}]$ – their required values, and by $Y_{it}$ – $i^{th}$ device's bandwidth resource. In our vision, the perfect machine type (MTC) rural 5G MAC protocol represents a MTC-enhanced flexible multifunctional distributed long-delay medium access control technology (MFMAC [8]), including also the functions of guaranteed dynamical control up to real time ("on the fly") of the bandwidth resources $\{Y_{it}\}$ allocation, guaranteed dynamical control accordingly to $k^{th}$data service classes of the traffic parameters $\{S_{ikt}\}$ and soft/different QoS parameters $\{X_{ikt}\}$, i.e., personally guaranteed Quality of Experience (QoE) for any user/device.

The required qualities of a perfect rural 5G MAC protocol may be implemented as MTC-aimed enhancements of the multifunctional distributed long-delay medium access control techniques [8, 12] exactly:

- high efficiency, tolerance, and lower latency [12], higher throughput and minimal overheads both come nearing fundamental limits [11] for a distributed multiple access control to long-delay wireless/space mediums;
- high controllability, reliability, stability, flexibility, and guarantee of distributed dynamical ("on the fly") control of broadband RAN technologies [8, 9, 12];
- multifunctional and universality abilities that rely on the dynamically controlled and adaptive ATM-like smart unified protocol MAC, i.e., MFMAC [8, 12], through the entire wireless networking hierarchy – core, backbone, and access networks;
- fully mesh all-device-centric radio access architecture all_device-to-all_device (D*m*D, *m*>>2) relies on the multipoint-to-multipoint (MPMP) [9] Virtual Space/Wireless ATM Hyperbus topology with fully distributed QoS-guaranteed multi-functional long-delay MAC [8];
- cost-effective completely distributed (grid-like) all-IP/MPLS over ATM-MFMAC Hyperbus (MFHBA) that implements the data packet selecting technique rather than packet switching technique [8, 9].

### 3.2 Fundamental Limits of the Distributed MAC

As showed [11], the real reachable throughput for various MAC protocols depends on ensuring the "MAC collective intellect" that contains a plenitude of information about the real-time state of the multiple access processes in geographically distributed queues, and also on the normalized overhead for provisioning QoS. What is the minimum reachable, or infimum, MAC overhead? And what is the potential reachable maximum throughput, or fundamental limit of potential capacity, of the ideal MAC protocol? It is reasonable to find the MAC overhead infimum as the Shannon entropy of the distributed multiple access processes based on the Markov models of distributed queues, and to find the potential capacity of MAC protocols as a function of the overhead infimum.

Let we define the real throughput capacity for real MAC protocol specified by real structural specifications and system parameters $\Gamma$, and by real medium conditions $\Psi$ including presence of errors as

$$C_{\Gamma,\Psi} = \max_{\{G \in F_G\}} S_{\Gamma,\Psi}(G), \tag{13}$$

where $F_G$ is the field of the possible values of input traffic intensity $G$. As in [11], we define the MAC throughput fundamental limit as supremum of the real throughput (13) on the set $F_\Gamma$ of MAC protocol's possible structural specifications and system parameters by given medium conditions $\Psi$, i.e., as potential capacity,

$$C_{\Psi}^{\sup} = \sup_{\{G \in F_G, \Gamma \in F_\Gamma\}} S_{\Gamma,\Psi}(G) = M[\tau]/(M[\tau] + \delta_{\Psi}^{\inf}) = 1/(1 + \nu_{\Psi}^{\inf}), \tag{14}$$

where $\delta_{\Psi}^{\inf}$ is the potential reachable minimum, or infimum, of the time resource overhead for medium access control per data unit/packet by duration $M[\tau]$,

$$\delta_{\Psi}^{\inf} = \inf_{\{\Gamma \in F_{\Gamma}\}} \delta_{\Gamma,\Psi}, \tag{15}$$

$v_{\Psi}^{\inf}$ is the normalized value of the infimum of overhead (15) according $M[\tau]$. The MACs overhead and throughput fundamental limits for widespread queueing models of distributed multiple access systems TDMA determine the statements [11]:

**Theorem 1:** If the TDMA system is described by an infinite model of equivalent centralized M/M/1 queue $\Gamma_0$ by $\Psi_0$ zero errors channel conditions, then the value of minimum reachable overhead on distributed MAC control is equal to

$$\inf_{\{\Gamma \in F_{\Gamma}\}} v_{\Gamma,\Psi_0} = v_{\Gamma_0,\Psi_0}^{\inf} = [2 + H(\tau)]/BM[\tau], \tag{16}$$

and the potential throughput capacity of the ideal MAC is equal to

$$\sup_{\{\Gamma \in F_{\Gamma}; G \in F_G\}} S_{\Psi_0,\Gamma}(G) = C_{\Psi_0}^{\sup} = 1/(1 + (2 + H(\tau))/BM[\tau]), \tag{17}$$

where $B$ is the bit rate, $M[\tau]$ is the mean duration of traffic packets, $H(\tau)$ is the entropy of the packets duration distribution given by the geometric law [11].

**Theorem 2:** If the TDMA system is described by the infinite model of equivalent centralized M/D/1 queue $\Gamma_0$ under conditions described in Theorem 1, then the value of minimum reachable expenses on distributed MAC control is equal to

$$\inf_{\{\Gamma \in F_{\Gamma}\}} v_{\Gamma,\Psi_0} = v_{\Gamma_0,\Psi_0}^{\inf} = 1,854/BM[\tau] \tag{18}$$

and the potential throughput capacity of the ideal MAC protocol is equal to

$$\sup_{\{\Gamma \in F_{\Gamma}; G \in F_G\}} S_{\Psi_0,\Gamma}(G) = C_{\Psi_0}^{\sup} = (1/(1 + 1,854/BM[\tau]). \tag{19}$$

We observe in Fig. 3, that the MAC's total entropy, i.e., overhead infimums (16) and (18), depends mainly from the data slots duration law indeterminacy. The M/M/1/* systems family must be characterized by greatest entropy in accordance with its exponential law's greatest indeterminacy. Opposite them, the M/D/1/* systems which are described by deterministic duration law ensure the least entropy, therefore – the greatest MAC potential throughput capacity (17) and (19). As proved in [11], the adaptive controlled multiple access MAC protocols with deterministic packet size and, hence – the least overheads, allow to reach to a fundamental limit of a MACs throughput capacity which, in turn, as much close to 1,0 as it's wished. The fully distributed ATM-like multifunctional MAC technology (MFMAC) [8, 12] meets the above breakthrough qualifications.

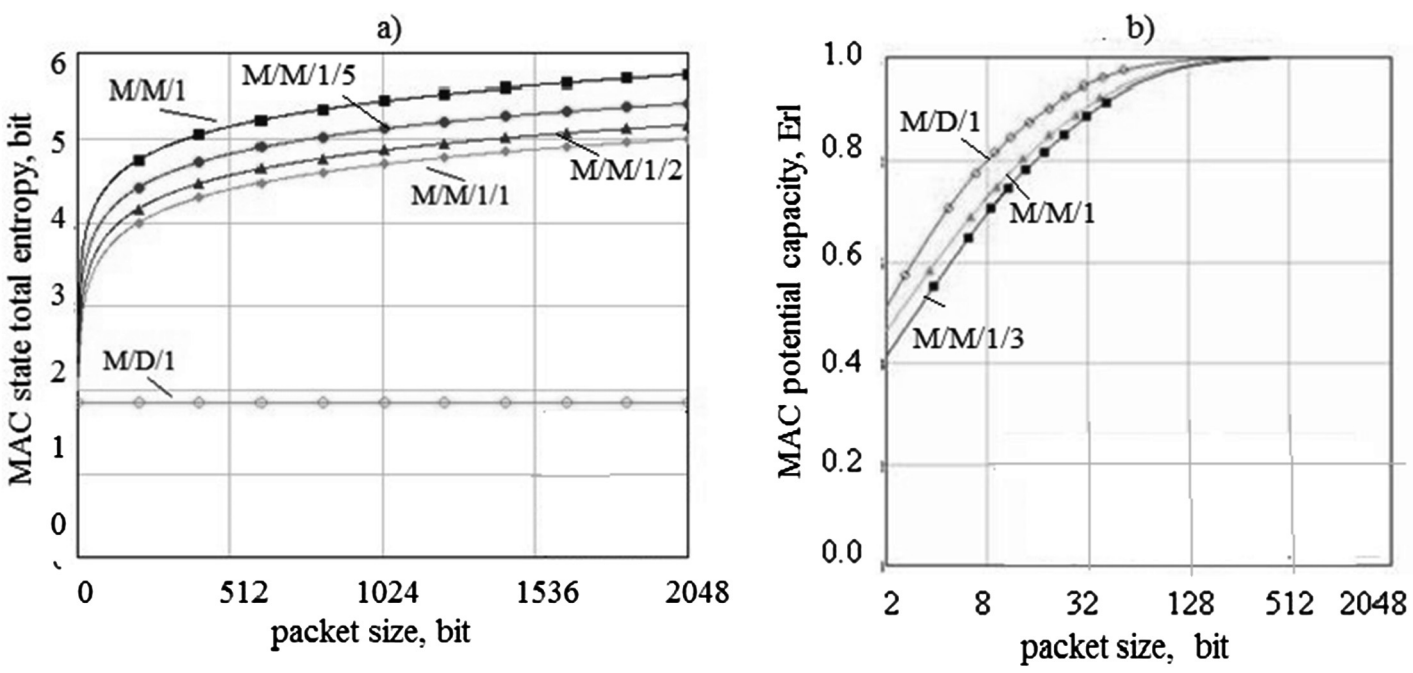

**Fig. 3.** MACs state entropy, or overhead infimum (a), and MACs throughput supreme (b) versus packets duration laws, SER = 1.0E-3.

The disruptive MFMAC technology uses the recurrent M-sequences (RS) MAC addressing opportunities [8, 14] in order to organize a RS-token tools "all-in-one" for high effective multiple access to long-delay space medium, soft QoS provision and distributed dynamical control of traffic parameters and bandwidth resources [8, 12, 14] approaching the sublimit of throughput capacity. The $M$-subsequences $A_j = a_{j-(n-1)}, a_{j-(n-2)}, \ldots, a_j$ serve as RS-identifiers of the unique MAC addresses and other protocol subjects. Some subset of $i^{\text{th}}$ "personal" identifiers $\{A^i_{jkt} | k = 1, 2, \ldots, m_{it}\} = B_{it}$ are dynamically assigned to each $i^{\text{th}}$ station on a decentralized basis by Shannon-Fano method for passing of the user's request in proportion to the required bandwidth resource $[Y_{it}]$.

## 4 Concept of Ubiquitous IoT/M2/H2H Green RRD 5G System

The RRDs extremely green device-centric Hypercelle is explained in Fig. 4. A conceptual look of the IP over DVB-2S multifunctional satellite-based fully distributed hybrid 5G networking technology RCS-MFMAC for RRD areas is explained in Fig. 5.

The hybrid architecture relies on implementation of the QoS-guaranteed multifunctional 5G machine type MAC perfect rural PHY-MAC techniques basing on the developing of the advanced delay-tolerant 5G ATM-like MPMP MFMAC technologies [8, 12] which in turn should be adapted to conditions of the satellite platforms' DVB-2S-RCS [10], VSAT, etc. The main breakthrough drivers for RRD-oriented 5G communications include also a push MFMAC-based next generations of wireless asynchronous transfer mode (ATM/MFMAC), of multi-protocol label switching (MPLS/MFMAC), and also of IP over DVD-S/MFMAC integrated networking technologies [9].

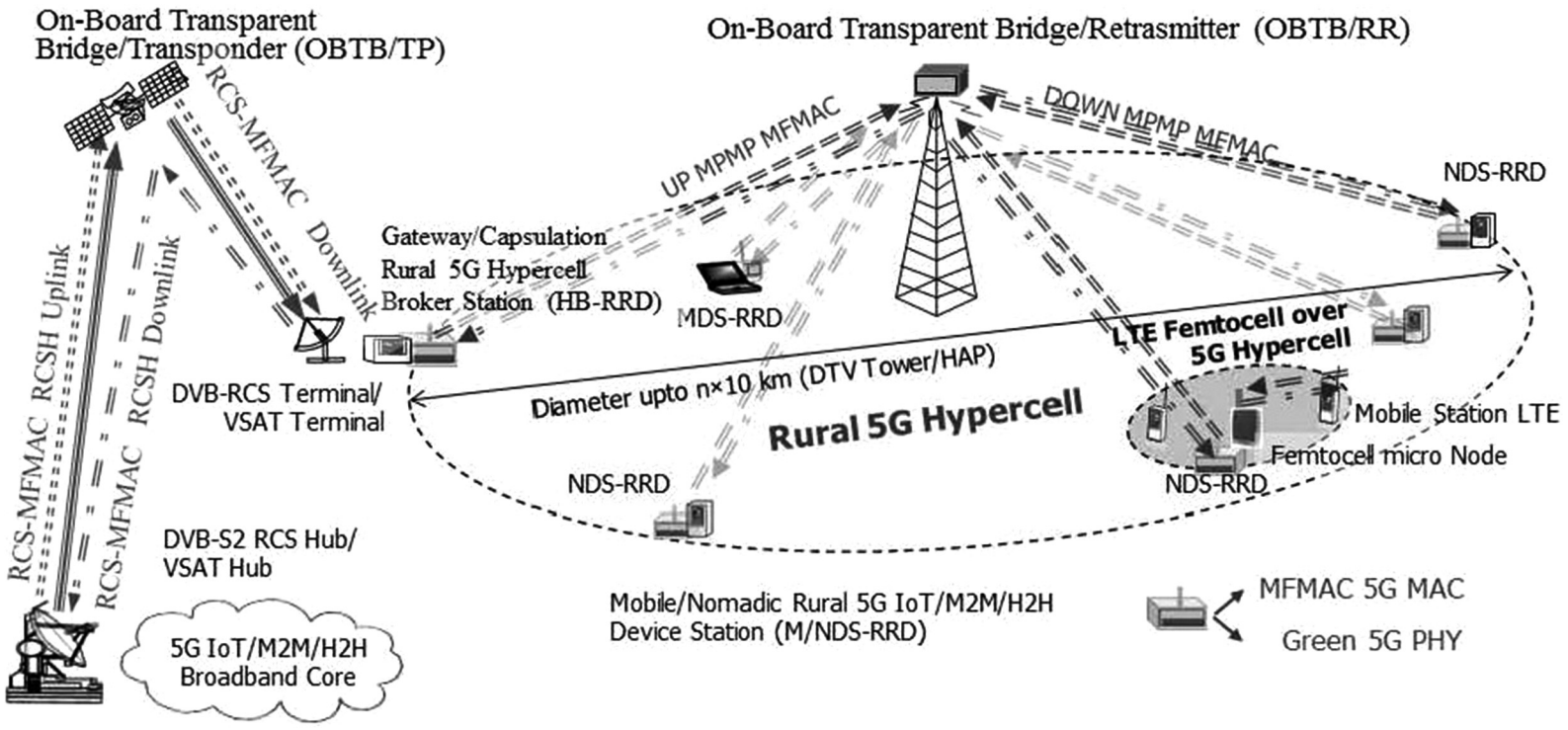

**Fig. 4.** Rural extremely green 5G Hypercelle.

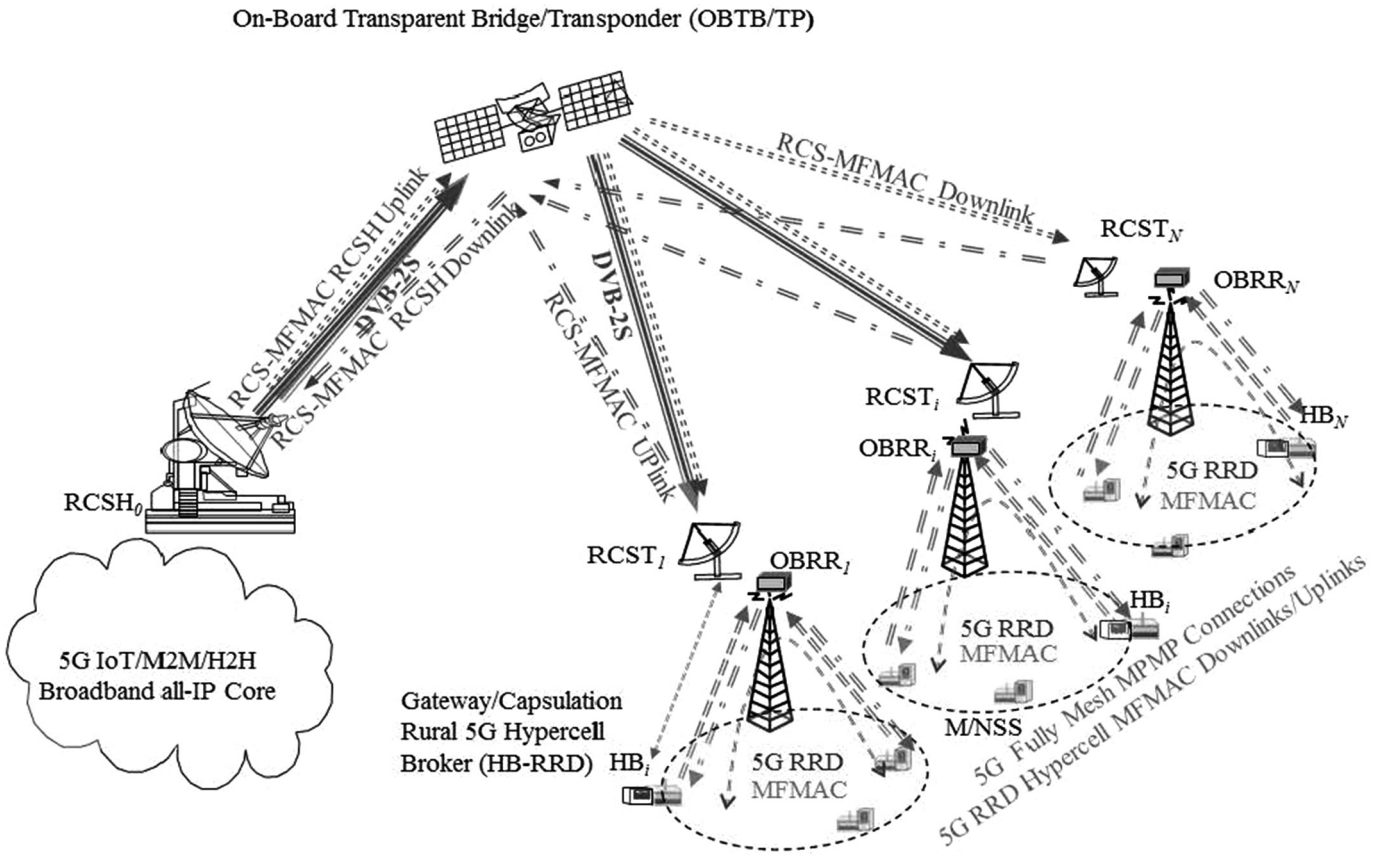

**Fig. 5.** Ubiquitous Green Rural 5G Hybrid Architecture.

## 5 Conclusion

In this paper, the green, ecological and cost-effective advanced approach to creation of the flexible QoS-guaranteed ubiquitous 5G IoT/M2M/H2H multifunctional RRD communications has been designed. Offered approach rely on implementation of the extremely flexible, energy and spectral effective 5G PHY-MAC techniques: (i) smarter increase of the SINR through beamforming/antenna/orthogonality gain, without rise of the transmitter power; (ii) closing "on-the-fly" the fundamental minimum of power consumption ICPE; (iii) providing "on-the-fly" the profitability/power efficiency aimed fundamental trade-offs for rural green 5G networks in practical invariant variables notions. It should be noted the key mission critical opportunities of a perfect rural 5G MAC: (j) the reachable low overhead close to fundamental infimum; (jj) the flexible 5G scheduler adapt "on-the-fly" the superframe formats and optimally allocate the massive machine type and also multiservice traffic by equal ATM-like minimal bandwidth block per second, without superframe overflow or redundancy.